\documentclass{article}
\usepackage[utf8]{inputenc}
\usepackage[T1]{fontenc}
\usepackage{lmodern}
\usepackage[margin=1in]{geometry}
\usepackage[colorlinks=true,linkcolor=blue,urlcolor=blue,citecolor=blue]{hyperref}

\usepackage{authblk}
\usepackage{amssymb}
\usepackage{amsfonts}
\usepackage{amsmath}

\begin{document}
	
\title{Inhomogeneous dust collapse in five-dimensional AdS-Chern-Simons gravity}

\author[1,2]{Luis Avil\'es\thanks{
		\href{mailto:luaviles@unap.cl}
		{\texttt{luaviles@unap.cl}}}}

\author[1,2]{Jos\'e D\'iaz\thanks{
		\href{mailto:joseludi@unap.cl }
		{\texttt{joseludi@unap.cl}}}}

\author[3]{Cristian Quinzacara\thanks{
		\href{mailto:cristian.quinzacara@uss.cl}
		{\texttt{cristian.quinzacara@uss.cl}}}}	

\author[1,2]{Patricio Salgado\thanks{
		\href{mailto:patsalgado@unap.cl}
		{\texttt{patsalgado@unap.cl}}}}	

\affil[1]{Facultad de Ciencias, Universidad Arturo Prat, Avda. Arturo Prat 2120, Iquique 1110939, Chile}
\affil[2]{Instituto de Ciencias Exactas y Naturales (ICEN), Universidad Arturo Prat,\newline Avda. Playa Brava 3256, Iquique 1110939, Chile}

\affil[3]{Departamento de Ciencias Exactas, Facultad de Ingeniería, Universidad San Sebastián, Concepción, Chile}

\date{}
\maketitle

\begin{abstract}
We study the gravitational collapse of a spherically symmetric, inhomogeneous dust cloud in five-dimensional AdS Chern--Simons gravity. Working in the torsion-free sector and using comoving Lema\^{\i}tre--Tolman--Bondi coordinates, we derive the reduced field equations and obtain an exact collapsing interior solution. The areal radius evolves periodically in time, but the physical collapsing branch extends only up to shell focusing or an earlier loss of regularity, while the initial density determines a radial function whose square is proportional to the conserved quasilocal mass. This separates the roles of the initial velocity and density profiles in controlling shell focusing and trapping.

We derive conditions for central regularity and for avoiding shell crossing, analyze the Kretschmann scalar, and show that shell focusing corresponds to a genuine curvature singularity, the weak and strong energy conditions are preserved throughout the regular collapsing phase. The homogeneous limit reduces consistently to a five-dimensional FLRW interior.

The collapsing cloud is matched smoothly to the black-hole branch of the static AdS Chern--Simons exterior. The null expansions show that a finite-radius apparent horizon forms only when the quasilocal mass exceeds a critical threshold. Initially untrapped shells above this threshold become trapped before shell focusing, whereas below the total-mass threshold no finite-radius future-trapped symmetry sphere forms in the regular interior. 
\end{abstract}

\section{Introduction}\label{sec:introduction}

Gravitational collapse provides one of the most direct settings in which the formation of spacetime singularities and trapped regions can be studied dynamically. Under appropriate causal and energy conditions, the singularity theorems imply geodesic incompleteness during sufficiently strong collapse \cite{haw}. They do not, however, determine whether the resulting singularity is hidden from distant observers. That question belongs to the cosmic censorship problem \cite{censura} and requires a separate analysis of the causal structure of the collapsing spacetime.

The classical model of Oppenheimer and Snyder \cite{oppen} describes the collapse of a homogeneous spherical dust cloud by matching a Friedmann interior to a Schwarzschild exterior. Relaxing homogeneity leads to the Lema\^{\i}tre--Tolman--Bondi (LTB) family \cite{lem,tolm,bondi}, in which the density and velocity profiles provide a richer set of initial data. Studies of spherical inhomogeneous dust collapse have shown that these data can determine whether the resulting singularity is hidden or visible \cite{cg10,cg11}, with related analyses extending to higher dimensions \cite{Ghosh2001,cg3} and quasispherical geometries \cite{cg2}. The range of collapse models has also been broadened through studies of Husain spacetimes \cite{cg1,cg14}, electromagnetic and scalar fields \cite{cg5}, and dark matter and dark energy \cite{cg6}. These different settings illustrate the importance of specifying both the matter content and the initial configuration. For the dust problem considered here, comoving coordinates allow one to follow individual shells, distinguish shell focusing from shell crossing \cite{joshi}, and compare their occurrence with the formation of trapped regions.

Modifications of the gravitational dynamics provide another source of departures from the classical dust-collapse picture, as illustrated by null-dust collapse in $f(R)$ gravity \cite{cg15}. Lovelock gravity is particularly suitable for studying higher-curvature effects because its field equations remain second order despite the inclusion of higher powers of the curvature \cite{Lovelock1971,Zumino1986}. Collapse in this theory has been investigated for spherical inhomogeneous dust with vanishing cosmological constant \cite{Ohashi2011} and for generalized Vaidya spacetimes \cite{cg4}. In five dimensions, the Lovelock action reduces to the Einstein--Gauss--Bonnet (EGB) action, whose vacuum branches and static spherical solutions already exhibit departures from Einstein gravity \cite{Boulware1986,Wheeler1986}. Dust collapse in this theory has been studied in spherical \cite{maeda,cg8} and quasispherical \cite{cg9} settings, revealing changes in the formation and causal character of singularities. 

More recent studies have examined the dependence of marginally trapped surfaces on the initial density and velocity profiles in five-dimensional EGB gravity \cite{Chatterjee2022}, as well as collapse in pure Gauss--Bonnet gravity for dust and more general fluid sources \cite{Dialektopoulos2023,Kumar2025}. Complementary analyses have considered spherical dust thin shells in EGB gravity \cite{Huang2022}. A recent analysis of Lovelock dust collapse has further related the highest curvature order to near-center trapping and the local visibility of shell-focusing singularities \cite{Ganguly2026}. Complementary work on homogeneous dust cosmologies has investigated the evolution equations in higher-dimensional EGB gravity with a cosmological constant \cite{Singh2025}. 

Within five-dimensional Einstein--Gauss--Bonnet gravity, the AdS Chern--Simons point is singled out by a special relation between the Gauss--Bonnet coupling and the cosmological constant. At this point, the two constant-curvature vacuum branches coalesce into a single AdS vacuum, and the theory admits a formulation as a Chern--Simons gauge theory for the AdS group \cite{cham,Tron2000}. Its static black-hole solutions belong to the dimensionally continued family \cite{Banados1994}, subsequently incorporated into the broader Lovelock classification of Cris\'ostomo, Troncoso, and Zanelli \cite{crisostomo}. Collapse in dimensionally continued gravity has been studied for homogeneous dust in odd dimensions \cite{Ilha1999} and for null dust \cite{Nozawa2006}. These studies provide the background for examining inhomogeneous timelike dust at the five-dimensional AdS Chern--Simons point, where the factorized field equations allow the shell dynamics and trapping conditions to be treated explicitly.

In this work we study the collapse of a spherically symmetric, inhomogeneous dust cloud in five-dimensional AdS Chern--Simons gravity. We restrict attention to spinless matter and to the torsion-free sector, and use comoving coordinates of LTB type. The initial data consist of a regular density profile and an inward velocity profile, and the evolution is followed through the regular collapsing region prior to shell crossing. Denoting by $R(t,r)$ the areal radius of the shell labeled by the comoving coordinate $r$, the reduced field equations at the Chern--Simons point separate in a particularly simple way: the shell evolution satisfies
\begin{equation}
	\ddot R+\frac{R}{l^2}=0,
\end{equation}
while the density determines a radial function $F(r)$ whose square is proportional to the conserved quasilocal mass. This makes it possible to obtain the interior solution explicitly and to track separately the roles played by the initial velocity and density profiles.

We then analyze the regularity conditions at the center, the avoidance of shell crossing, and the curvature behavior of the resulting solution. The Kretschmann scalar distinguishes the regular center from shell-focusing and shell-crossing singularities, and the homogeneous limit reduces consistently to a five-dimensional FLRW interior. The collapsing cloud is matched across a comoving boundary to the black-hole branch of the static AdS Chern--Simons exterior. This choice fixes the corresponding positive branch of the interior function $F$ through the matching conditions and relates the quasilocal mass of the cloud to the exterior mass parameter.

Particular attention is given to the formation of trapped surfaces. The radial null expansions yield a finite-radius apparent horizon only for shells satisfying $F(r)>1$, or equivalently for quasilocal mass above a definite threshold. For initially untrapped configurations above this threshold, the apparent horizon is crossed before shell focusing, while below the total-mass threshold no finite-radius future-trapped symmetry sphere forms in the regular interior. Regularity nevertheless forces $F(r)\to0$ toward the center, so the innermost shells reach focusing without first encountering a finite-radius apparent horizon. This observation concerns the trapping structure only; it does not by itself determine whether the central singularity is locally or globally visible. Establishing that requires a separate analysis of outgoing null geodesics, which is left for future work.

The paper is organized as follows. Section~\ref{sec:action} introduces the five-dimensional Einstein--Gauss--Bonnet action and the AdS Chern--Simons point. Section~\ref{sec:reduction} reduces the field equations for a spherically symmetric dust source and shows that the angular equation is redundant in the regular nondegenerate sector. Section~\ref{sec:solution} constructs the interior solution, relates the initial data to the quasilocal mass, and examines regularity, curvature, energy conditions, and the homogeneous limit. Section~\ref{sec:matching} matches the collapsing cloud to the black-hole branch of the static AdS Chern--Simons exterior. Section~\ref{sec:horizons} studies the null expansions, apparent horizons, and the associated mass threshold. The main results and their limitations are summarized in Section~\ref{sec:conclusions}.

\section{The gravitational action and the Chern--Simons point}
\label{sec:action}

Lovelock's theorem \cite{Lovelock1971,Zumino1986} states that the most general action in 5 dimensions, which does not include torsion and which leads to second-order field equations for the metric, is given by a linear combination of the Einstein--Hilbert term, a cosmological term, and a particular quadratic curvature term known as the Gauss--Bonnet term. The resulting action is the Einstein--Gauss--Bonnet action. When this gravitational action is coupled to a matter action $S_{\rm m}$ and treated in the first-order formalism, with torsion not necessarily vanishing, it reads
\begin{equation}\label{eq:egb-action}
	S[e,\omega]=\frac{1}{12\kappa_5}\int_M\varepsilon_{abcde}
	\left(\alpha R^{ab}R^{cd}e^e+R^{ab}e^ce^de^e
	-\frac{\Lambda}{10}e^ae^be^ce^de^e\right)+S_{\rm m}.
\end{equation}
The independent gravitational fields are the \textit{vielbein} (coframe) $e^a=e^a{}_{\mu}dx^\mu$ and the spin connection $\omega^{ab}$. The vielbein determines the spacetime metric through $g_{\mu\nu}=\eta_{ab}e^a{}_{\mu}e^b{}_{\nu}$, while the spin connection defines the curvature two-form $R^{ab}=d\omega^{ab}+\omega^a{}_{c}\omega^{cb}$.
The constant $\alpha$ is the Gauss--Bonnet coupling, $\Lambda$ is the cosmological constant, and $\kappa_5$ is the five-dimensional gravitational coupling. Throughout, wedge products between differential forms are understood implicitly. We adopt the signature $(-++++)$.

For general EGB gravity, $\alpha$ and $\Lambda$ are independent parameters,
with dimensions of length squared and inverse length squared, respectively. Setting
$\alpha=0$ leaves Einstein gravity with cosmological constant. In this work,
however, we choose these parameters in terms of a single length scale $l$,
\begin{equation}
	\alpha=\frac{3}{2}l^2,\qquad \Lambda=-\frac{3}{l^2}.
	\label{eq:cs-ads}
\end{equation}
These are the AdS Chern--Simons couplings in the normalization of \eqref{eq:egb-action} \cite{cham,Tron2000,crisostomo}. The length $l$ will be the curvature radius of the vacuum.

\subsection{AdS Chern--Simons point}

The special choice of couplings in \eqref{eq:cs-ads} can be understood as follows. Consider a maximally symmetric vacuum of EGB gravity satisfying $R^{ab}=\lambda e^ae^b$, where $\lambda$ is a constant to be determined. Substituting this \textit{ansatz} into the field equations obtained from \eqref{eq:egb-action} gives
\begin{equation}
	P(\lambda)\equiv\alpha\lambda^2+3\lambda-\frac{\Lambda}{2}=0.
	\label{eq:vacuum-polynomial}
\end{equation}
In general, this quadratic equation may admit two distinct vacuum curvatures. For the couplings \eqref{eq:cs-ads}, however, it becomes
\begin{equation}
	P(\lambda)=\frac{3l^2}{2}
	\left(\lambda+\frac{1}{l^2}\right)^2,
	\qquad \lambda=-\frac{1}{l^2}.
	\label{eq:ads-double-root}
\end{equation}
The two roots therefore coincide, leaving a single constant-curvature vacuum: AdS with curvature radius $l$.

This is not the only choice of couplings for which \eqref{eq:vacuum-polynomial} factorizes in the form of \eqref{eq:ads-double-root}. The other possibility is the dS Chern--Simons choice $\alpha=-3l^2/2$ and $\Lambda=+3/l^2$, which gives the double root $\lambda=+1/l^2$. We restrict the rest of the paper to the AdS case.

Once \eqref{eq:cs-ads} is imposed, $\alpha$ is no longer independent of $\Lambda$. In particular, taking $\alpha\to0$ while preserving this relation sends $\Lambda\to-\infty$. Therefore, this family does not admit Einstein gravity with fixed cosmological constant as a limit.

\subsection{Field equations in the torsion-free sector}

The matter source considered in this work is spinless dust. Its vanishing spin current makes the connection field equation obtained from \eqref{eq:egb-action} homogeneous in the torsion:
\begin{equation}
	\varepsilon_{abcde}\mathcal T^c
	\left(2\alpha R^{de}+3e^de^e\right)=0,
	\qquad \mathcal T^a=de^a+\omega^a{}_{b}e^b.
	\label{eq:connection}
\end{equation}
Equation \eqref{eq:connection} admits $\mathcal T^a=0$. We choose this sector, so the spin connection becomes the Levi--Civita connection of the metric. This allows us to rewrite the action \eqref{eq:egb-action} in the familiar metric form
\begin{equation}
	S[g]=\frac{1}{2\kappa_5}\int_M d^5x\sqrt{-g}
	\left(R-2\Lambda+\frac{\alpha}{6}\mathcal L_{\rm GB}\right)+S_{\rm m},
	\label{eq:metric-action}
\end{equation}
where the Gauss--Bonnet contribution is
\begin{equation}
	\mathcal L_{\rm GB}=R^2-4R_{\mu\nu}R^{\mu\nu}
	+R_{\mu\nu\rho\sigma}R^{\mu\nu\rho\sigma}.
	\label{eq:gb-density}
\end{equation}
The factor \(\alpha/6\) follows from the normalization of the first-order action \eqref{eq:egb-action} and fixes our convention for the Gauss--Bonnet coupling.

The coupling to matter is described through the usual energy-momentum tensor
\begin{equation}
	T_{\mu\nu}=-\frac{2}{\sqrt{-g}}\frac{\delta S_{\rm m}}{\delta g^{\mu\nu}}.
\end{equation}
Varying \eqref{eq:metric-action} with respect to the metric then gives
\begin{equation}
	\mathcal E_{\mu\nu}\equiv G_{\mu\nu}
	+\frac{\alpha}{6}H_{\mu\nu}
	+\Lambda g_{\mu\nu}
	=\kappa_5 T_{\mu\nu},
	\label{eq:field-equations}
\end{equation}
where $G_{\mu\nu}$ is the Einstein tensor and $H_{\mu\nu}$ contains the quadratic curvature contribution,
\begin{equation}
	H_{\mu\nu}=2\left(RR_{\mu\nu}-2R_{\mu\rho}R_\nu{}^\rho
	-2R^{\rho\sigma}R_{\mu\rho\nu\sigma}
	+R_{\mu\rho\sigma\tau}R_\nu{}^{\rho\sigma\tau}\right)
	-\frac12g_{\mu\nu}\mathcal L_{\rm GB}.
	\label{eq:lanczos-tensor}
\end{equation}
Thus, $\mathcal E_{\mu\nu}$ collects the complete geometric side of the field equations: the Einstein contribution, the Gauss--Bonnet correction, and the cosmological term. Equation \eqref{eq:field-equations} remains valid for arbitrary EGB couplings. From the next section onward, however, we specialize to the AdS Chern--Simons values \eqref{eq:cs-ads}.


\section{Spherically symmetric dust: reduction of the field equations}
\label{sec:reduction}

For a spherically symmetric, spinless dust cloud, we use comoving proper time and write \cite{Ghosh2001,maeda}
\begin{equation}
	ds^2=-dt^2+A^2(t,r)\,dr^2+R^2(t,r)\,d\Omega_3^2,
	\qquad T_{\mu\nu}=\rho(t,r) u_\mu u_\nu,
	\qquad u^\mu=\delta^\mu_t.
	\label{eq:dust-metric}
\end{equation}
Here $d\Omega_3^2$ is the metric of the unit three-sphere, $R(t,r)$ is the \emph{areal} radius, and dots and primes denote derivatives with respect to $t$ and $r$, respectively. We work in a regular region with $A>0$ and $R>0$, while the center is understood as a limiting point.

At the AdS Chern--Simons point defined in \eqref{eq:cs-ads}, the field equations take a particularly compact form. Before writing them down, it is convenient to introduce the following functions:
\begin{subequations}
	\begin{align}
		F(t,r)&=\dot R^2+1-\frac{R'^2}{A^2}+\frac{R^2}{l^2},
		\label{eq:aux-F}\\
		G(t,r)&=\frac{2A'R'}{A^3}-\frac{2R''}{A^2}
		+\frac{2\dot A\dot R}{A}+\frac{2R}{l^2}.
		\label{eq:aux-G}
	\end{align}
\end{subequations}
Substituting the metric \eqref{eq:dust-metric} into \eqref{eq:field-equations}, the temporal, mixed, radial, and angular components reduce to
\begin{subequations}
	\begin{align}
		\mathcal E^t{}_t&=-\frac{3l^2}{2R^3}FG=-\kappa_5\rho,
		\label{eq:dust-tt}\\
		\mathcal E^t{}_r&=\frac{3l^2A}{R^3}F\,
		\partial_t\!\left(\frac{R'}{A}\right)=0,
		\label{eq:dust-tr}\\
		\mathcal E^r{}_r&=-\frac{3l^2}{R^3}F
		\left(\ddot R+\frac{R}{l^2}\right)=0,
		\label{eq:dust-rr}\\
		\mathcal E^i{}_i&=-\frac{l^2}{R^2}\left[
		F\left(\frac{\ddot A}{A}+\frac{1}{l^2}\right)
		+G\left(\ddot R+\frac{R}{l^2}\right)
		+2\left(\partial_t\left(\frac{R'}{A}\right)\right)^2
		\right]=0.
		\label{eq:dust-ii}
	\end{align}
\end{subequations}
These equations contain the basic dynamics of the dust cloud in the Chern--Simons theory. The temporal equation determines the matter density, while the remaining equations constrain the evolution of the geometry.

We now consider the angular equation \eqref{eq:dust-ii}. The geometric tensor $\mathcal E^\mu{}_\nu$ satisfies the divergence identity
\begin{equation}
	\nabla_\mu\mathcal E^\mu{}_\nu=0,
\end{equation}
which is the Lovelock counterpart of the contracted Bianchi identity. Spherical symmetry restricts the geometric tensor to equal angular components, while imposing the mixed equation $\mathcal E^t{}_r=0$ makes it diagonal. For the metric \eqref{eq:dust-metric}, the radial component of the divergence identity then becomes
\begin{equation}
	0=\nabla_\mu\mathcal E^\mu{}_r
	=\partial_r\mathcal E^r{}_r+
	\frac{3R'}{R}\left(\mathcal E^r{}_r-\mathcal E^i{}_i\right),
	\label{eq:radial-bianchi}
\end{equation}
where $i$ denotes any angular direction, without summation. The radial field equation gives $\mathcal E^r{}_r=0$ and therefore also $\partial_r\mathcal E^r{}_r=0$. Hence, whenever $R\ne0$ and $R'\ne0$, equation \eqref{eq:radial-bianchi} implies
\begin{equation}
	\mathcal E^i{}_i=0.
\end{equation}
Thus, in the dust sector the angular equation is not independent because it follows from the mixed and radial equations together with the divergence identity.

\section{Interior solution and admissible initial data}
\label{sec:solution}

In a region containing matter with $\rho>0$, equation \eqref{eq:dust-tt} requires $F\neq0$. The mixed and radial equations \eqref{eq:dust-tr} and \eqref{eq:dust-rr} then give
\begin{equation}
	R'=A\, W(r),
	\qquad
	\ddot R+\frac{R}{l^2}=0,
	\label{eq:ltb-sector}
\end{equation}
where $W(r)$ is an arbitrary function of the comoving radial coordinate. We restrict attention to the region before shell crossing, where $R'>0$. The temporal equation \eqref{eq:dust-tt} then determines the matter density.

To specify the evolution, we choose an initial hypersurface $t=0$ in the regular collapsing region. The residual freedom in the choice of radial coordinate can be fixed by labeling each dust shell with its initial areal radius,
\begin{equation}
	R(0,r)=r.
	\label{eq:initial-radius}
\end{equation}
We then prescribe the initial inward velocity of each shell as
\begin{equation}
	\dot R(0,r)=-v(r),
	\qquad
	v(r)\ge0,
	\label{eq:initial-velocity}
\end{equation}
where regularity at the center requires $v(0)=0$ and, for a smooth center, $v(r)$ must vanish at least linearly with $r$. Together with $W(r)$, the function $v(r)$ specifies the initial configuration of the dust cloud. The relation of $W(r)$ to the initial density and mass profiles will be established in Subsection \ref{subsec:density}.

With these initial conditions, the solution of \eqref{eq:ltb-sector} can be written as
\begin{equation}
	R(t,r)=l\,h(r)\sin\left(\frac{1}{l}\left(t_c(r)-t\right)\right),
	\label{eq:collapse-solution}
\end{equation}
where
\begin{equation}
	h(r)=\sqrt{v^2(r)+\frac{r^2}{l^2}},
	\qquad
	t_c(r)=l\arctan\left(\frac{r}{l\,v(r)}\right),
	\label{eq:amplitude-phase}
\end{equation}
for $v(r)>0$, while $t_c(r)=\pi l/2$ when $v(r)=0$. The function $h(r)$ fixes the amplitude of the motion, while $t_c(r)$ is the proper time at which the shell labeled by $r$ reaches zero areal radius.

Thus, during $0<t<t_c(r)$, each shell has $R>0$ and $\dot R<0$ and moves monotonically inward. The evolution is followed only up to shell focusing, $R=0$, or to any earlier loss of regularity such as shell crossing.

\subsection{Density and radial integration functions}\label{subsec:density}

In order to solve the temporal field equation \eqref{eq:dust-tt}, we first collect some useful consequences of the previous results for the functions $F(t,r)$ and $G(t,r)$.

First, multiplying the radial equation in \eqref{eq:ltb-sector} by $2\dot R$ allows it to be written as a conserved quantity,
\begin{equation}
	\frac{\partial}{\partial t}\left(\dot R^2+\frac{R^2}{l^2}\right)=0
	\quad\Longrightarrow\quad
	\dot R^2+\frac{R^2}{l^2}=h^2(r),
	\label{eq:first-integral}
\end{equation}
where $h(r)$ is the same function introduced in \eqref{eq:amplitude-phase}, as can be verified by substituting the solution \eqref{eq:collapse-solution}.

Equation \eqref{eq:first-integral}, together with $R'=AW(r)$ from \eqref{eq:ltb-sector}, reduces the definition \eqref{eq:aux-F} to
\begin{equation}
	F(r)=1+h^2(r)-W^2(r).
	\label{eq:F-radial}
\end{equation}
Thus, $F$ depends only on the comoving radial coordinate and is therefore constant along each dust shell.

Second, a useful relation between $F$ and $G$ follows directly from their definitions. Differentiating $F$ with respect to $r$ using \eqref{eq:aux-F} and comparing the result with $R'G$, obtained from \eqref{eq:aux-G}, gives
\begin{equation}
	F'=R'G+2A\dot R\,\partial_t\left(\frac{R'}{A}\right)=R'G,
	\label{eq:Fprime-G}
\end{equation}
where in the second equality we use the mixed field equation \eqref{eq:dust-tr}.

Using this identity, the temporal field equation \eqref{eq:dust-tt} becomes
\begin{equation}
	\frac{3l^2}{4R^3R'}(F^2)'=\kappa_5\rho.
	\label{eq:density-integral}
\end{equation}
This equation shows that two types of singular behavior may arise: shell-crossing singularities, characterized by ($R'=0$), and shell-focusing singularities, characterized by ($R=0$).
We now evaluate this equation on the initial hypersurface $t=0$. Since the radial coordinate was fixed by $R(0,r)=r$, it follows directly that $R'(0,r)=1$. Denoting the initial density profile by $\rho_0(r)=\rho(0,r)$, equation \eqref{eq:density-integral} becomes
\begin{equation}\label{eq:f2prime}
	(F^2)'=
	\frac{4\kappa_5}{3l^2}\rho_0(r)r^3.
\end{equation}
This equation relates $F(r)$ directly to the initial density profile of the cloud. Furthermore, because $F$ depends only on the shell label $r$, equation \eqref{eq:density-integral} at an arbitrary time can be compared with its value at $t=0$, giving the density in terms of the metric function $R$,
\begin{equation}
	\rho(t,r)
	=
	\frac{\rho_0(r)r^3}{R^3R'}.
	\label{eq:density-evolution}
\end{equation}

Returning to \eqref{eq:f2prime}, its radial integration introduces the constant $F(0)$, which is fixed by regularity at the center. From $R'=AW(r)$ and the initial gauge $R(0,r)=r$, we have $R'(0,0)=1$. Local regularity of the radial metric \eqref{eq:dust-metric} then requires $A(0,0)=1$, so that
\begin{equation}
	W(0)=1.
	\label{eq:winitial}
\end{equation}
Regularity of the initial velocity also gives $v(0)=0$, and therefore $h(0)=0$ from \eqref{eq:amplitude-phase}. Using \eqref{eq:F-radial}, we finally obtain
\begin{equation}
	F(0)=1+h^2(0)-W^2(0)=0.
\end{equation}
Using this condition, integration of \eqref{eq:f2prime} from the center to the shell $r$ yields
\begin{equation}
	F^2(r)=\frac{4\kappa_5}{3l^2}
	\int_0^r \rho_0(s)s^3\,ds.
	\label{eq:initial-mass-data}
\end{equation}

Thus, the initial density profile determines $F^2(r)$ through \eqref{eq:initial-mass-data}, while the initial velocity determines $h(r)$. The function $F$ is therefore determined only up to a sign. For a continuous solution with nonzero density, the sign of $F$ is fixed on each connected radial region. Equation \eqref{eq:F-radial} then determines
\begin{equation}
	W^2(r)=1+h^2(r)-F(r),
\end{equation}
subject to the admissibility condition $W^2>0$. Since $W(0)=1$, we choose the continuous branch $W>0$ in the regular region. Before shell crossing, where $R'>0$, the remaining metric function is therefore
\begin{equation}
	A=\frac{R'}{W}.
\end{equation}

At this stage, the interior solution is completely determined once the initial profiles $\rho_0(r)$ and $v(r)$ are specified together with a choice of branch for $F$. Equation \eqref{eq:initial-mass-data} fixes $F^2(r)$ from the density, while \eqref{eq:amplitude-phase} fixes $h(r)$ and $t_c(r)$ from the velocity. The remaining radial function then follows from
\begin{equation}
	W^2(r)=1+h^2(r)-F(r),
\end{equation}
and $A=R'/W$ determines the radial metric coefficient before shell crossing.

\subsection{Quasilocal mass}

The function $F(r)$ introduced above can be related directly to the mass contained inside each comoving shell. For a spherically symmetric spacetime, the generalized Misner--Sharp mass in five-dimensional EGB gravity, with the normalization used here, reduces at the AdS Chern--Simons point to \cite{MaedaNozawa2008}
\begin{equation}
	m(r)=\frac{3\pi^2l^2}{2\kappa_5}F^2(r).
	\label{eq:quasilocal-mass}
\end{equation}
Thus, the quantity $F^2(r)$ obtained from the temporal field equation in \eqref{eq:initial-mass-data} is proportional to the quasilocal mass enclosed by the shell labeled by $r$.

Combining \eqref{eq:quasilocal-mass} with \eqref{eq:density-integral} gives
\begin{equation}
	m'=2\pi^2\rho R^3R',
	\qquad
	\dot m=0.
	\label{eq:mass-balance}
\end{equation}
The first relation shows how the enclosed mass changes from one shell to the next, while the second expresses conservation of mass along each comoving dust shell. Evaluating the first relation on the initial hypersurface and using $m(0)=0$ gives
\begin{equation}
	m(r)=2\pi^2\int_0^r\rho_0(s)s^3\,ds.
	\label{eq:mass-integral}
\end{equation}
The condition $m(0)=0$ follows from central regularity and is consistent with $F(0)=0$.

This quasilocal mass should not be confused with the proper-volume integral of the rest-mass density, whose radial volume element is $2\pi^2AR^3\,dr$. The relation between $m(r)$ and the mass parameter of the exterior solution will be established through the matching conditions.

\subsection{Central regularity and shell crossing}\label{subsec:regularity}

We now impose the conditions required for the interior solution to remain regular near the center and free of shell-crossing singularities during the collapse.

For a smooth center, the initial density must approach a finite value without developing a preferred radial direction. We therefore write
\begin{equation}\label{eq:density}
	\rho_0(r)=\rho_c+O(r^2).
\end{equation}
Similarly, the initial velocity must vanish at the center and, for a smooth radial profile, has the form
\begin{equation}\label{eq:velocity}
	v(r)=v_1r+O(r^3).
\end{equation}
Using \eqref{eq:initial-mass-data}, the first condition implies $F^2(r)=O(r^4)$ and therefore $F(r)=O(r^2)$. On the other hand, from \eqref{eq:amplitude-phase} we have $h^2(r)=O(r^2)$, which through equation \eqref{eq:F-radial} gives $W(r)=1+O(r^2)$, where the positive branch has been chosen consistently with the regular-center condition $W(0)=1$.

A second possible loss of regularity occurs when neighboring dust shells cross each other. In comoving coordinates, this happens when $R'=0$ while $R>0$. To exclude shell crossing, we therefore require $R'>0$ throughout the regular collapsing phase.

For this purpose, it is useful to rewrite \eqref{eq:collapse-solution} using the initial conditions:
\begin{equation}
	R(t,r)=r\cos\left(\frac{t}{l}\right)
	-lv(r)\sin\left(\frac{t}{l}\right).
\end{equation}
Differentiating with respect to $r$ and rearranging gives
\begin{equation}
	R'=\cos\left(\frac{t}{l}\right)
	-lv'(r)\sin\left(\frac{t}{l}\right)
	=\frac{R}{r}
	+l\left(\frac{v}{r}-v'\right)\sin\left(\frac{t}{l}\right).
	\label{eq:no-crossing}
\end{equation}
Before focusing, $R>0$ and $0\leq t<t_c\leq\pi l/2$, so $\sin(t/l)\geq0$. Hence, $v'\leq v/r$ guarantees $R'>0$. Conversely, if $v'>v/r$, the right-hand side of \eqref{eq:no-crossing} approaches a negative value as $t\to t_c(r)^-$. Since $R'$ is continuous in time and initially $R'=1$, it must pass through zero at some time $t_{\rm sc}<t_c(r)$. At that time the shell still has $R>0$, so a shell crossing occurs before focusing.

The necessary and sufficient condition for avoiding shell crossing throughout each shell's collapsing interval is therefore
\begin{equation}
	v'(r)\leq\frac{v(r)}{r},
	\qquad r>0.
	\label{eq:cond}
\end{equation}
Since
\begin{equation}
	t_c'(r)=\frac{v(r)-rv'(r)}{h^2(r)},
\end{equation}
this is equivalent to $t_c'(r)\geq0$: the focusing time must not decrease outward.

\subsection{Curvature and singularities}

We now examine whether the loss of regularity identified above corresponds to a genuine curvature singularity. For $A=R'/W$, contraction of the Riemann tensor gives the Kretschmann scalar
\begin{equation}
	K\equiv R_{\mu\nu\rho\sigma}R^{\mu\nu\rho\sigma}
	=4\frac{\ddot R'^2}{R'^2}+12\frac{\ddot R^2}{R^2}
	+12\frac{(\dot R\dot R'-WW')^2}{R^2R'^2}
	+12\frac{(\dot R^2+1-W^2)^2}{R^4}.
	\label{eq:kretschmann-geometric}
\end{equation}
Using \eqref{eq:ltb-sector} together with the fact that $F=F(r)$, this expression reduces to
\begin{equation}
	K=\frac{16}{l^4}
	+12\left(\frac{F'}{2RR'}-\frac{1}{l^2}\right)^2
	+12\left(\frac{F}{R^2}-\frac{1}{l^2}\right)^2.
	\label{eq:kretschmann-solution}
\end{equation}
For a fixed noncentral shell with $F(r)\neq0$, the last term in \eqref{eq:kretschmann-solution} diverges as $R\to0$. Shell focusing therefore corresponds to a scalar curvature singularity. Likewise, if $R'=0$ while $R>0$ and $F'\neq0$, the second term diverges, showing that a shell-crossing event is also curvature singular in this case.

The center, however, requires a separate limiting analysis. Since $R(t,0)=0$ throughout the evolution, the vanishing of the areal radius at $r=0$ does not by itself signal a singularity. For smooth initial data with the central density $\rho_c>0$ defined in \eqref{eq:density}, we may write
\begin{equation}
	F(r)=f_2r^2+O(r^4),
	\qquad
	f_2^2:=\frac{\kappa_5\rho_c}{3l^2}.
\end{equation}
Likewise, substituting the central behavior $v(r)=v_1r+O(r^3)$ introduced in \eqref{eq:velocity} into \eqref{eq:amplitude-phase} gives
\begin{equation}
	h(r)=h_1r+O(r^3),
	\qquad
	t_c(r)=t_{c0}+O(r^2),
\end{equation}
where the corresponding leading coefficients are
\begin{equation}
	h_1:=\sqrt{v_1^2+\frac{1}{l^2}},
	\qquad
	t_{c0}=l\arctan\left(\frac{1}{lv_1}\right).
\end{equation}
The solution \eqref{eq:collapse-solution} therefore behaves near the center as
\begin{equation}
	R(t,r)=a(t)r+O(r^3),
	\qquad
	a(t)=lh_1\sin\left(\frac{t_{c0}-t}{l}\right).
\end{equation}
Taking the limit $r\to0$ in \eqref{eq:kretschmann-solution} then gives
\begin{equation}
	K(t,0)=\frac{16}{l^4}
	+24\left(\frac{f_2}{a^2(t)}-\frac{1}{l^2}\right)^2.
	\label{eq:central-curvature}
\end{equation}
The central curvature therefore remains finite as long as $a(t)>0$ and diverges when $a(t)\to0$, which marks central shell focusing. Thus, the condition $R=0$ at the symmetry center does not by itself signal a singularity; what matters is the limiting behavior of the curvature.

\subsection{An explicit inhomogeneous collapse model}

To illustrate the previous results, consider the initial density profile
\begin{equation}
	\rho_0(r)=\rho_c\left[1-\left(\frac{r}{r_b}\right)^n\right],
	\qquad
	0\leq r\leq r_c\leq r_b,
	\label{eq:explicit-density}
\end{equation}
where $\rho_c>0$ is the central density, $r_b$ is the radial scale at which the extrapolated density profile vanishes, $r_c$ is the comoving radius of the physical boundary of the cloud, and $n$ is a positive even integer controlling how rapidly the density decreases outward. This restriction on $n$ ensures that the density is smooth at the center, consistent with the regularity assumptions of Subsection~\ref{subsec:regularity}. The physical cloud therefore occupies $0\leq r\leq r_c$, and its density need not vanish at the matching surface.

Substituting \eqref{eq:explicit-density} into \eqref{eq:initial-mass-data} gives
\begin{equation}
	F^2(r)=\frac{\kappa_5\rho_c}{3l^2}\,
	r^4\left[
	1-\frac{4r^n}{(n+4)r_b^n}
	\right].
	\label{eq:explicit-F}
\end{equation}
The corresponding quasilocal mass follows immediately from \eqref{eq:quasilocal-mass},
\begin{equation}
	m(r)=\frac{\pi^2\rho_c}{2}\,
	r^4\left[
	1-\frac{4r^n}{(n+4)r_b^n}
	\right].
	\label{eq:explicit-mass}
\end{equation}
As expected, the explicit dependence on $l$ and $\kappa_5$ cancels from the mass profile. The gravitational couplings determine how the matter distribution enters the geometry through $F(r)$, while the mass enclosed by a given shell is fixed directly by the initial matter distribution.

The expressions \eqref{eq:explicit-F} and \eqref{eq:explicit-mass} apply throughout the physical cloud, $0\leq r\leq r_c$. The parameter $r_b$ sets the radial scale of the chosen density profile, while the matching surface is located independently at $r=r_c$. In particular, a smooth matching to the vacuum exterior does not require $\rho_0(r_c)=0$.

To complete the initial data, choose the inward velocity profile
\begin{equation}
	v(r)=v_1r,
	\qquad v_1>0.
\end{equation}
The amplitude and focusing time are then
\begin{equation}
	h(r)=h_1r,
	\qquad
	h_1=\sqrt{v_1^2+\frac{1}{l^2}},
	\qquad
	t_c(r)=t_{c0}
	=l\arctan\left(\frac{1}{lv_1}\right).
\end{equation}
Thus, all shells focus simultaneously, and the areal radius takes the form
\begin{equation}
	R(t,r)=a(t)r,
	\qquad
	a(t)=lh_1\sin\left(\frac{t_{c0}-t}{l}\right).
\end{equation}
Since $R'=a(t)>0$ for $0\leq t<t_{c0}$, no shell crossing occurs before focusing. The density evolves as
\begin{equation}
	\rho(t,r)=\frac{\rho_c}{a^4(t)}
	\left[1-\left(\frac{r}{r_b}\right)^n\right].
\end{equation}
Although all shells share the same contraction factor, the density remains radially inhomogeneous.

On the positive-$F$ branch, the remaining metric functions are
\begin{equation}
	W^2(r)=1+h_1^2r^2-F(r),
	\qquad
	A(t,r)=\frac{a(t)}{W(r)},
\end{equation}
where $F(r)$ is the positive square root of \eqref{eq:explicit-F}. An initially untrapped configuration requires
\begin{equation}
	F(r)<1+\frac{r^2}{l^2},
	\qquad 0<r\leq r_c,
\end{equation}
which also guarantees $W^2>0$. A simple sufficient parameter restriction is
\begin{equation}
	\kappa_5\rho_c l^2\leq3.
\end{equation}
Indeed, \eqref{eq:explicit-F} then gives $F(r)\leq r^2/l^2$, ensuring the preceding inequality throughout the cloud. Together with $0<r_c\leq r_b$ and positive even $n$, these choices provide an explicit family of regular, initially untrapped configurations without shell crossing before focusing.

\subsection{Energy conditions}

For pressureless dust, the weak and strong energy conditions both reduce to the requirement
\begin{equation}
	\rho\geq0.
\end{equation}
Indeed, the dust stress tensor has vanishing radial and tangential pressures, while in five dimensions the strong energy condition requires $\rho+p\geq0$ and $2\rho+4p\geq0$, which are both satisfied whenever $\rho\geq0$.

For the profile \eqref{eq:explicit-density}, $\rho_0(r)\geq0$ throughout $0\leq r\leq r_c$ provided $\rho_c>0$. Moreover, from \eqref{eq:density-evolution},
\begin{equation}
	\rho(t,r)=\frac{\rho_0(r)r^3}{R^3R'}.
\end{equation}
Hence, as long as the evolution remains in the regular collapsing region with $R>0$ and $R'>0$, the density remains nonnegative. The weak and strong energy conditions are therefore preserved throughout the collapse up to shell focusing or an earlier shell-crossing singularity.

\subsection{Homogeneous limit}

As a consistency check, consider
\begin{equation}
	\rho_0(r)=\rho_c,
	\qquad
	v(r)=v_1r.
\end{equation}
Then \eqref{eq:amplitude-phase} gives
\begin{equation}
	R(t,r)=a(t)r,
	\qquad
	a(t)=lh_1\sin\left(\frac{t_{c0}-t}{l}\right),
\end{equation}
with $h_1^2=v_1^2+l^{-2}$ and $t_{c0}=l\arctan\big((lv_1)^{-1}\big)$.

For constant density, \eqref{eq:initial-mass-data} gives $F(r)=f_2r^2$, where
\begin{equation}
	f_2=\sqrt{\frac{\kappa_5\rho_c}{3l^2}}.
\end{equation}
Hence, from \eqref{eq:F-radial},
\begin{equation}
	W^2(r)=1-kr^2,
	\qquad
	k:=f_2-h_1^2.
\end{equation}
Since $A=R'/W$, the metric reduces to
\begin{equation}
	ds^2=-dt^2+a^2(t)
	\left[
	\frac{dr^2}{1-kr^2}
	+r^2d\Omega_3^2
	\right].
\end{equation}
Thus, the homogeneous sector of the solution is the five-dimensional FLRW geometry.\\

The interior solution is therefore specified by two independent pieces of initial data. The density profile $\rho_0(r)$ determines $F^2(r)$ and the quasilocal mass, while the velocity profile $v(r)$ determines the shell amplitude $h(r)$ and the collapse time $t_c(r)$. Once the branch of $F$ is fixed, $W(r)$ follows from \eqref{eq:F-radial}, and the metric function $A$ is obtained from $A=R'/W$. The remaining task is to determine which interior branch is compatible with the black-hole branch of the static exterior metric and how the interior mass is related to the exterior mass parameter.

\section{Matching to the AdS Chern--Simons black-hole}
\label{sec:matching}

We now match the collapsing cloud to the black-hole branch of the static, spherically symmetric vacuum solution of five-dimensional AdS Chern--Simons gravity found in \cite{crisostomo}. In the notation of Crisóstomo, Troncoso, and Zanelli, the five-dimensional Chern--Simons theory corresponds to $d=5$ and $k=2$. Their AdS radius is the same length $l$ used throughout this work, while their gravitational constant $G_2$ is related to our coupling $\kappa_5$ by
\begin{equation}
	G_2=\frac{\kappa_5}{3\pi^2l^2}.
	\label{eq:G2-kappa5}
\end{equation}
With these conventions, the exterior geometry can be written as
\begin{equation}
	ds_+^2=-f(x)dT^2+\frac{dx^2}{f(x)}+x^2d\Omega_3^2,
	\qquad
	f(x)=1+\frac{x^2}{l^2}-\sqrt{1+2G_2M}.
	\label{eq:exterior}
\end{equation}
Here $x$ is the exterior areal radius and $M$ is the mass parameter used in \cite{crisostomo}. Notice that this parameter is shifted with respect to the AdS vacuum: pure AdS corresponds to
\begin{equation}
	M=-\frac{1}{2G_2},
\end{equation}
whereas $M=0$ gives the zero-radius black-hole limit. We take the branch with the negative square root in \eqref{eq:exterior}, which admits a positive-radius black-hole horizon for $M>0$. As we will show, this choice selects the corresponding branch of the interior function $F$ through the matching conditions.

The surface of the cloud is the comoving timelike hypersurface
\begin{equation}
	\Sigma:\quad r=r_c.
\end{equation}
Using the proper time $\tau$ of the boundary, its trajectory is described from the interior by
\begin{equation}
	R_\Sigma(\tau)=R(\tau,r_c),
\end{equation}
and from the exterior by
\begin{equation}
	x=R_\Sigma(\tau),
	\qquad
	T=T_\Sigma(\tau).
\end{equation}
The induced metric is therefore the same on both sides,
\begin{equation}
	ds_\Sigma^2=-d\tau^2+R_\Sigma^2(\tau)d\Omega_3^2.
\end{equation}

To obtain a smooth matching without a surface layer, we impose continuity of the extrinsic curvature across $\Sigma$. This is a sufficient condition for satisfying the Einstein--Gauss--Bonnet junction conditions in the absence of a surface stress tensor \cite{Davis2003}. With the unit normal directed from the interior toward the exterior, define
\begin{equation}
	\beta=\sqrt{\left(\partial_\tau R_\Sigma\right)^2+f(R_\Sigma)}.
	\label{eq:beta}
\end{equation}
The independent nonvanishing components of the extrinsic curvature are then
\begin{equation}
	K^i{}_{i-}=\frac{W_c}{R_\Sigma},
	\qquad
	K^\tau{}_{\tau-}=0,
	\qquad
	K^i{}_{i+}=\frac{\beta}{R_\Sigma},
	\qquad
	K^\tau{}_{\tau+}
	=
	\frac{\partial_\tau^2 R_\Sigma+\tfrac12\partial_x f(R_\Sigma)}{\beta},
	\label{eq:extrinsic-curvatures}
\end{equation}
where $W_c=W(r_c)>0$ and there is no sum over $i$.

\subsection{Matching the evolution and mass}

Continuity of the angular component of the extrinsic curvature across the boundary gives
\begin{equation}
	K^i{}_{i-}=K^i{}_{i+}
	\qquad\Longrightarrow\qquad
	\beta=W_c.
\end{equation}
Using the definition of $\beta$ given in \eqref{eq:beta}, this condition becomes
\begin{equation}
	\left(\partial_\tau R_\Sigma\right)^2+\frac{R_\Sigma^2}{l^2}
	=
	W_c^2-1+\sqrt{1+2G_2M}.
	\label{eq:boundary-evolution}
\end{equation}
On the other hand, the interior first integral \eqref{eq:first-integral} evaluated at the boundary gives
\begin{equation}
	\left(\partial_\tau R_\Sigma\right)^2+\frac{R_\Sigma^2}{l^2}
	=
	h_c^2,
\end{equation}
where $h_c=h(r_c)$. Since \eqref{eq:F-radial} implies
\begin{equation}
	F(r_c)=1+h_c^2-W_c^2,
\end{equation}
comparison with \eqref{eq:boundary-evolution} yields
\begin{equation}
	F(r_c)=\sqrt{1+2G_2M}.
	\label{eq:F-boundary}
\end{equation}
Matching to the chosen exterior solution therefore selects the positive branch of $F$ at the boundary. For a continuous nonvanishing $F$ in the connected interior region, this fixes the positive branch throughout that region.

Squaring \eqref{eq:F-boundary} gives
\begin{equation}
	M=\frac{F^2(r_c)-1}{2G_2}.
	\label{eq:mass-matching}
\end{equation}
Using the coupling relation \eqref{eq:G2-kappa5} together with the quasilocal mass \eqref{eq:quasilocal-mass},
\begin{equation}
	m(r_c)=\frac{F^2(r_c)}{2G_2},
\end{equation}
and therefore
\begin{equation}
	m(r_c)=M+\frac{1}{2G_2}.
	\label{eq:mass-offset}
\end{equation}
Thus, the interior quasilocal mass is the exterior mass measured relative to the regular AdS vacuum. The additive term $1/(2G_2)$ reflects the convention used in \cite{crisostomo}, for which pure AdS corresponds to $M=-1/(2G_2)$.

The temporal components of the extrinsic curvature are automatically continuous. Indeed, for the exterior solution,
\begin{equation}
	\frac{1}{2}\partial_x f(R_\Sigma)=\frac{R_\Sigma}{l^2},
\end{equation}
while the interior evolution gives
\begin{equation}
	\partial_\tau^2 R_\Sigma=-\frac{R_\Sigma}{l^2}.
\end{equation}
Hence $K^\tau{}_{\tau+}=0=K^\tau{}_{\tau-}$. This equality is obtained without dividing by $\partial_\tau R_\Sigma$ and therefore remains valid also at a turning point.

\section{Trapped spheres and apparent horizons}
\label{sec:horizons}
As the cloud contracts, an important question is whether outgoing light rays can still increase the areal radius of the symmetry spheres. This is determined by the expansions of the future-directed radial null congruences. We choose
\begin{equation}
	k_\pm=\partial_t\pm A^{-1}\partial_r,
	\qquad
	k_+\cdot k_-=-2,
\end{equation}
for which the corresponding expansions are
\begin{equation}
	\theta_\pm
	=
	\frac{3}{R}k_\pm(R)
	=
	\frac{3}{R}\left(\dot R\pm W\right).
	\label{eq:null-expansions}
\end{equation}
During collapse, $\dot R<0$ and $W>0$, so the ingoing expansion is always negative, $\theta_-<0$. The onset of trapping is therefore controlled by the outgoing expansion. A marginally outer trapped sphere is reached when $\theta_+=0$, namely when
\begin{equation}
	\dot R=-W.
\end{equation}
Equivalently,
\begin{equation}
	g^{\mu\nu}\partial_\mu R\partial_\nu R
	=
	W^2-\dot R^2
	=
	1+\frac{R^2}{l^2}-F(r)
	=
	0.
	\label{eq:trapping-condition}
\end{equation}
The apparent-horizon radius is therefore
\begin{equation}
	R_{\rm AH}^2(r)
	=
	l^2\big(F(r)-1\big).
	\label{eq:apparent-radius}
\end{equation}
This result does not require $W=1$. A regular marginally trapped sphere can exist only for $F(r)>1$. For such a shell, spheres with
\begin{equation}
	R<R_{\rm AH}(r)
\end{equation}
are future trapped. If $F(r)\leq1$, no marginally trapped or future-trapped symmetry sphere occurs at finite $R>0$.

\subsection{Initial absence of trapped surfaces}

Before studying when trapped surfaces form during the collapse, we require that they are not already present on the initial hypersurface. From \eqref{eq:null-expansions}, the outgoing expansion at $t=0$ is
\begin{equation}
	\theta_+(0,r)
	=
	\frac{3}{r}\big(W(r)-v(r)\big).
\end{equation}
Thus, an initially untrapped shell satisfies
\begin{equation}
	W(r)>v(r).
\end{equation}
Using $h^2(r)=v^2(r)+r^2/l^2$ together with \eqref{eq:F-radial}, this condition can be written entirely in terms of $F(r)$ as
\begin{equation}
	F(r)<1+\frac{r^2}{l^2},
	\qquad
	0<r\leq r_c.
	\label{eq:initial-untrapped}
\end{equation}
Equivalently,
\begin{equation}
	g^{\mu\nu}\partial_\mu R\partial_\nu R
	\bigg|_{t=0}
	=
	1+\frac{r^2}{l^2}-F(r)
	>0.
\end{equation}
For shells with $F(r)\leq1$, condition \eqref{eq:initial-untrapped} is automatically satisfied. When $F(r)>1$, it becomes
\begin{equation}
	r>l\sqrt{F(r)-1}
	=
	R_{\rm AH}(r),
\end{equation}
so the initial areal radius of the shell lies outside its apparent-horizon radius. At the cloud boundary, using \eqref{eq:F-boundary} and \eqref{eq:exterior}, the same condition becomes
\begin{equation}
	f(r_c)>0.
\end{equation}
Thus, when the matched exterior admits a horizon, the boundary of an initially untrapped cloud is initially located outside it.

\subsection{Crossing time and mass threshold}

For shells with $F(r)>1$, the apparent horizon exists at finite radius. If such a shell is initially untrapped, condition \eqref{eq:initial-untrapped} places it outside this radius at $t=0$. As the collapse proceeds, the shell eventually reaches the marginally trapped surface. Using the solution \eqref{eq:collapse-solution} together with \eqref{eq:apparent-radius}, the corresponding crossing time is
\begin{equation}
	t_{\rm AH}(r)
	=
	t_c(r)
	-l\arcsin\left(
	\frac{\sqrt{F(r)-1}}{h(r)}
	\right).
	\label{eq:apparent-time}
\end{equation}

The difference between this time and the shell-focusing time provides a direct measure of how long the shell evolves inside the trapped region before reaching zero areal radius. Defining
\begin{equation}
	\Delta t(r)
	:=
	t_c(r)-t_{\rm AH}(r)
	=
	l\arcsin\left(
	\frac{\sqrt{F(r)-1}}{h(r)}
	\right),
	\label{eq:trapping-time-lag}
\end{equation}
we have $\Delta t(r)>0$ whenever $F(r)>1$. At fixed nonzero $h$, the interval $\Delta t$ tends to zero as $F\to1^+$ and increases with $F$. Comparisons between different shells, however, depend on both $F(r)$ and $h(r)$.

The argument of the inverse sine is well defined on the collapsing branch. Indeed,
\begin{equation}
	F(r)-1=h^2(r)-W^2(r),
\end{equation}
and therefore, for $F(r)>1$ and $W(r)>0$,
\begin{equation}
	0<
	\frac{F(r)-1}{h^2(r)}
	<1.
\end{equation}
For an initially untrapped shell, this implies
\begin{equation}
	0<t_{\rm AH}(r)<t_c(r).
\end{equation}
The shell therefore crosses the apparent horizon after the initial hypersurface but before shell focusing. The limiting case
\begin{equation}
	r=R_{\rm AH}(r)
\end{equation}
corresponds to $t_{\rm AH}(r)=0$, while $r<R_{\rm AH}(r)$ describes a shell that is already trapped on the initial hypersurface.

The condition for trapping can also be expressed directly in terms of the quasilocal mass. On the positive-$F$ branch selected by the matching conditions, $F(r)>1$ is equivalent, through \eqref{eq:quasilocal-mass}, to
\begin{equation}
	m(r)>m_*,
	\qquad
	m_*:=\frac{3\pi^2l^2}{2\kappa_5}
	=
	\frac{1}{2G_2}.
	\label{eq:mass-threshold}
\end{equation}
The scale $m_*$ therefore separates shells that can develop a finite-radius apparent horizon from those that cannot.

At the surface of the cloud, \eqref{eq:mass-offset} turns the same condition into
\begin{equation}
	M>0.
\end{equation}
The interior apparent horizon also joins continuously to the horizon of the matched static exterior, since
\begin{equation}
	R_{\rm AH}^2(r_c)
	=
	x_h^2
	=
	l^2\left(
	\sqrt{1+2G_2M}-1
	\right).
	\label{eq:exterior-horizon}
\end{equation}
Thus, the trapping threshold obtained from the interior dynamics is the same threshold that controls the existence of a positive-radius horizon in the exterior geometry.

Finally, for nonnegative density, the quasilocal mass $m(r)$ is nondecreasing with $r$. Hence, if
\begin{equation}
	m(r_c)\leq m_*,
\end{equation}
then $F(r)\leq1$ throughout the regular interior and no future-trapped symmetry sphere forms at finite $R>0$. In the limiting case $m(r_c)=m_*$, the apparent-horizon radius shrinks to zero at the boundary.

\subsection{Scope of the causal conclusions}

Regular central data satisfy $F(r)=O(r^2)$, so $F(r)<1$ for sufficiently small $r$. The innermost shells therefore reach shell focusing without crossing a finite-radius apparent horizon. By contrast, outer shells with sufficiently large $F(r)$ may become trapped before reaching their focusing time.

The causal behavior of the central singularity is determined by outgoing radial null curves, which satisfy
\begin{equation}
	\frac{dt}{dr}
	=
	A(t,r)
	=
	\frac{R'(t,r)}{W(r)},
	\qquad
	\frac{dR}{dt}
	=
	\dot R+W.
	\label{eq:outgoing-null}
\end{equation}
If an outgoing null curve emerges from the central singularity into the regular interior, the singularity is locally visible. Global visibility further depends on whether such a curve can cross the boundary of the cloud and propagate through the exterior spacetime toward the AdS conformal boundary.

The results obtained above therefore determine where trapped surfaces form, when each shell crosses the apparent horizon, and the mass threshold associated with their formation. A detailed analysis of the outgoing null trajectories, and hence of the local and global visibility of the central singularity, is left for future work.

\section{Discussion and conclusions}\label{sec:conclusions}

We have constructed the torsion-free LTB dust solution at the five-dimensional AdS Chern--Simons point. The special choice of Lovelock couplings factorizes the field equations and reduces the shell dynamics to $\ddot R+R/l^2=0$, while the radial Bianchi identity renders the angular equation redundant in the regular nondegenerate sector. The resulting contracting solution is explicit. Its formal continuation is periodic, but the physical evolution ends once shell focusing or an earlier loss of regularity is reached. As expected for the Chern--Simons theory, the coupling relation does not allow an Einstein limit at fixed cosmological constant.

One of the clearest consequences of this reduction is that the two initial profiles play markedly different roles. The velocity profile $v(r)$ fixes the amplitude of each shell and the time at which it reaches $R=0$. The density profile $\rho_0(r)$ instead fixes $F^2(r)$, and therefore the quasilocal mass. Once the branch of $F$ is selected, the density and velocity profiles together determine the radial function $W(r)$. The finite-radius trapping threshold is controlled by $F(r)$, whereas the shell-focusing time is controlled directly by $v(r)$. In the homogeneous limit this structure reduces consistently to a five-dimensional FLRW interior.

The curvature analysis separates the regular center from the genuine singularities of the collapse. Shell focusing produces a scalar curvature singularity, and shell crossing is also curvature singular whenever $F'\neq0$. For smooth initial data with positive central density, the curvature at the center remains finite throughout the regular phase and diverges at the central focusing time. For nonnegative initial density, the weak and strong energy conditions are preserved as long as $R>0$ and $R'>0$, while the condition obtained for the velocity profile prevents shell crossing before focusing. The conserved quasilocal mass is proportional to $F^2$, and the matching to the black-hole branch of the static AdS Chern--Simons exterior requires the positive branch of $F$ and yields
\begin{equation}
	m(r_c)=M+\frac{1}{2G_2},
	\qquad
	G_2=\frac{\kappa_5}{3\pi^2l^2}.
\end{equation}
The trapping structure is especially simple. Marginal symmetry spheres satisfy
\begin{equation}
	R_{\rm AH}^2(r)=l^2\big(F(r)-1\big),
\end{equation}
so a finite-radius apparent horizon can occur only when $F(r)>1$, or equivalently when the quasilocal mass exceeds $1/(2G_2)$. At the boundary this becomes $M>0$, and the interior apparent horizon joins continuously to the horizon of the static exterior. For initially untrapped shells with $F(r)>1$, trapping precedes shell focusing by the interval given in \eqref{eq:trapping-time-lag}. Conversely, if the total quasilocal mass satisfies $m(r_c)\leq1/(2G_2)$, no finite-radius future-trapped symmetry sphere develops anywhere in the regular interior.

These results determine when and where trapping occurs, but they do not decide whether the central singularity is visible. Regularity forces $F(r)\to0$ toward the center, so the innermost shells reach focusing without first encountering a finite-radius apparent horizon. Whether an outgoing null geodesic can emerge from that endpoint, propagate across the cloud boundary, and reach the AdS conformal boundary is a separate causal question. The local and global visibility of the singularity, together with its dependence on the admissible initial profiles, will be addressed elsewhere.

\subsection*{Acknowledgements}
L.A. was partially supported by ANID through FONDECYT
grant No.~11261098 and SIA-ANID grant No.~85220027; C.Q. was supported in
part by ANID-Chile through FONDECYT Grant $11231238$; P.S. was supported in
part by ANID-Chile through FONDECYT Grant $N^{o}$ $1262414$ and in part by
UNAP-VRII grant $N^{o}$ 091/25. The authors wish to thank Stephanie
Caro, Diego Molina, Sebastian Salgado, Cristian Vera, for enlightening discussions.

\end{document}